\documentclass[a4paper,11pt]{article}
\pdfoutput=1 

\usepackage{jinstpub} 

\usepackage{lineno}
\usepackage{subfig}

\title{\boldmath 
Online control of the gain drift with temperature of SiPM arrays  used for  the readout of LaBr$_3$:Ce crystals}

\author[a,b,1]{M.Bonesini,\note{Corresponding author.}}
\author[a]{R.Bertoni,}
\author[c]{M.Prata}
\author[c]{and M.Rossella}

\affiliation[a]{Sezione INFN Milano Bicocca,Piazza Scienza 3, Milano, Italy}
\affiliation[b]{Dipartimento di Fisica G. Occhialini, Universit\'a di Milano Bicocca, Piazza Scienza 3, Milano,Italy}
\affiliation[c]{Sezione INFN Pavia,via A. Bassi 6, Pavia, Italy}

\emailAdd{maurizio.bonesini@mib.infn.it}

\abstract{
LaBr$_3$:Ce crystals have been introduced for radiation imaging in 
medical physics, with photomultiplier or single SiPM readout. 
An R$\&$D was pursued with 1/2" and 1" LaBr$_3$:Ce crystals, from different producers, to realize 
compact large area detectors (up to some cm$^2$ area) with SiPM array readout, 
aiming at high light yields, good energy resolution, good detector linearity 
and fast time response for low-energy X-rays. A natural application was 
found inside the FAMU project at RIKEN-RAL muon facility, that aims at a precise 
measure of the proton Zemach radius to solve the so-called "proton radius 
puzzle", triggered by the recent measure of the proton charge radius at PSI. 
The goal is the detection of characteristic X-rays around 130 keV. Other 
applications may be foreseen in medical physics, such as PET, and gamma-ray 
astronomy. A limiting factor is the gain drift of SiPM arrays with temperature, 
that give a deterioration of the detector's FWHM   energy resolution. 
To solve this problem, a custom NIM module, based on CAEN A7585 digital power
supply, was developed.
Test results of the correction of gain drift with temperature
for SiPM arrays from Advansid, Sensl, Hamamatsu will be presented.
At the$^{137}$ Cs peak,  an energy resolution  better 
than  $ 3 \%$ was obtained for a typical LaBr$_3$:Ce crystal, using 
Hamamatsu S13461 arrays. This  compares well with best available results 
obtained with a PMT. 
}

\keywords{SiPMTs; X-ray detectors.}

\proceeding{12$^{\text{th}}$ International Conference on Position Sensitive Detectors\\
12-17 September, 2021\\
University of Birmingham, UK}

\begin{document}
\maketitle
\flushbottom

\section{Introduction}
\label{sec:intro}
\begin{figure}[htbp] 
\centering
\vskip -1.40 cm
\includegraphics[width=0.70\textwidth]{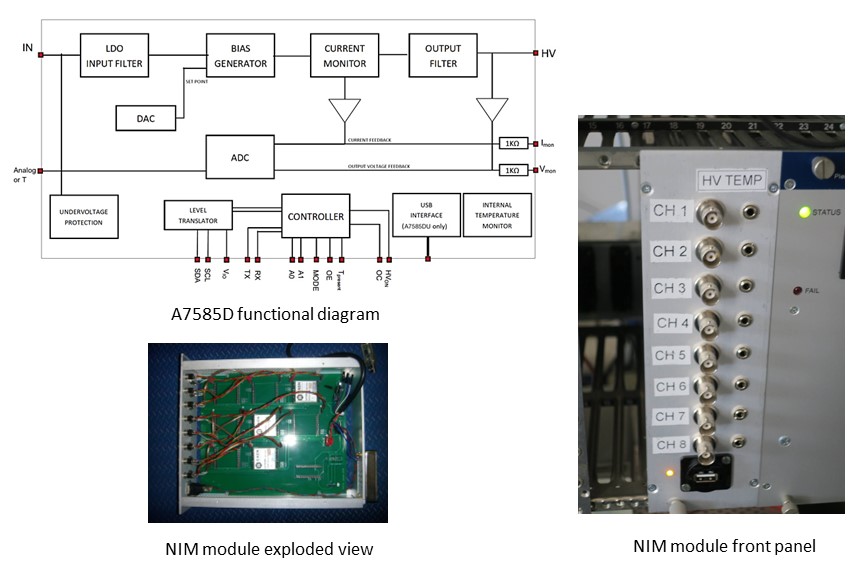}
\vskip -0.5cm
\caption{Left panel - top, A7585D functional layout (courtesy of CAEN srl);
        bottom, exploded view of one custom NIM module.
	Right panel: front panel of one custom NIM modules: a) is the BNC 
	 connector for a channel HV, b) the 3.5 mm jack stereo for the 
	 connection to a temperature TMP37 sensor, c) the USB interface 
	 connection.}
\label{fig-module}
\end{figure}

Silicon photomultipliers (SiPM) are a valuable alternative to conventional
photomultipliers (PMTs) for the readout of scintillation detectors. 
They may be arranged as an array. 
As readout devices they have a high reliability, a low sensitivity to external
magnetic fields and can operate at voltages significantly lower than the ones
used for PMTs. Using a SiPM array it is possible to obtain energy resolutions 
comparable to what obtained with PMTs. However, SIPMs have a relevant problem: in addition to an increased noise level, their gain drifts significantly as a function of temperature. 
This feature prevents their use in conditions with a changing 
temperature, as homeland security and military 
applications.
To overcome this problem, SiPMs must be operated at a fixed gain if 
temperature changes. Thus the operating voltage $V_{op}=V_{bd} + V_{ov}$,
with $V_{bd}$ breakdown voltage and $V_{ov}$ overvoltage, must be modified
as a function of temperature according to:
$$ V_{bd}(T) = V_{bd}(T_{0}) \times (1 + \beta (T-T_{0}) $$
with T working temperature, $T_{0}$ reference temperature (typically 25$^{\circ}$ C) 
and $\beta =\Delta V_{bd}/ \Delta T$ temperature coefficient of the used SiPM,
as explained in reference \cite{dinu}. 
For an online hardware correction, 
a custom NIM module was developed, based on CAEN A7585D digital
power supplies, with temperature feedback. Up to eight channels may be powered
by a single 2-slots NIM module, as shown in figure \ref{fig-module}.
The control of this module was implemented  via either
a FDTI USB-I2C converter or an  Arduino Nano chip. 
Three modules (with which up to 24 channels may be powered) were realized and are
linkable in daisy chain, via the I2C protocol.  
Our application has developed inside the FAMU experiment at RAL, to measure
with high precision the proton Zemach radius \cite{famu}: its main 
requirement was the measure of X-rays around 100 keV with compact crystal 
detectors. 
\section{Laboratory tests}
\label{sec:lab-tests}
\begin{figure}[htbp] 
\centering
\includegraphics[width=0.37\textwidth]{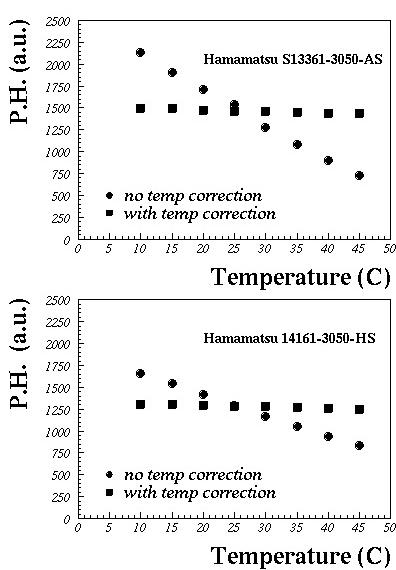}
\includegraphics[width=0.37\textwidth]{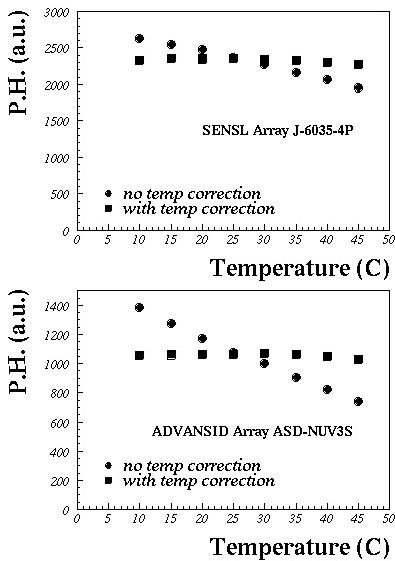}
\caption{Dependence of the photo peak position at 662 keV for a typical 1/2" 
detector, with and without temperature correction. Different SiPM arrays 
have been used for the readout.
}
\label{fig-crys}
\end{figure}

The crystal detectors are mounted inside an ABS holder, realized with a 3D printer and
are fully described in reference \cite{detec}. The SiPM array temperature is
measured by a TMP37 temperature sensor, mounted on the PCB where the array
socket is soldered. A 3.5 mm stereo jack cable connects the sensor to the 
custom NIM module for online temperature correction. 
Laboratory tests were done putting the detectors under test
inside a Memmert IPV-30 climatic chamber, where
the temperature could be stabilized with a precision of $\sim$ 0.1 $^{\circ}$C.
Detectors were
powered at their
nominal operating voltage $V_{op}$.
The summed analogue signal from the cells of a SiPM array is directly fed into a CAEN V1730
fast digitizer (500 MHz bandwidth, 14 bit resolution) and is acquired 
by a custom DAQ system. 
In the reported laboratory tests, a $^{137}$Cs exempt source was used and the
dependence of the photo peak at 662 keV, as a function of temperature is 
reported in figure \ref{fig-crys}. The same crystal is used in all the tests.
While Advansid and Hamamatsu SiPM arrays show a larger variation of response
(up to 70 $\%$), the effect is smaller for SENSL devices ($\sim 30 \%$).
After correction, in all cases the variation is reduced to $\sim 5 \%$. 

\begin{figure}[htbp] 
\centering
\vskip -0.2 cm
\includegraphics[width=0.80\textwidth]{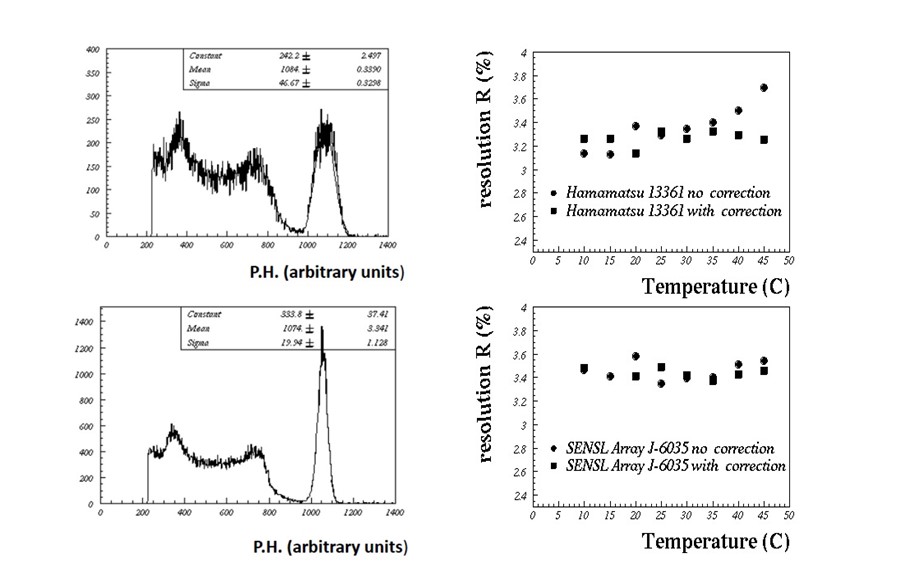}
	\caption{Left panel: Cs$^{137}$ spectra recorded by a LaBr$_{3}$:Ce 
	1/2" 
	detector read by an Advansid ASD-NUV3S SiPM  array during 
	a temperature scan between 20 $^{\circ}$C and
	30 $^{\circ}$C. Top (bottom) 
	is without (with) temperature correction. Fit results to the 662
	keV photo peak, with a simple gaussian, are reported.
	Right panel: dependence of the energy resolution at 662 keV on
	temperature, for a typical 1/2" LaBr$_3:$Ce detector read by different 
	SiPM arrays, with/without temperature correction.}
\label{fig:crys1}
\end{figure}

The response of a typical 1/2" crystal, inside the climatic chamber, 
with a temperature
excursion from 20 $^{\circ}$C to 30 $^{\circ}$C, 
is shown in the left panel of figure \ref{fig:crys1}. Without
the online correction, the resolution of the photo peak at 662 keV is 
sensibly degraded. The variation of FWHM energy resolution of the same
crystal, with different readout, is shown instead in the right panel.

\section{Conclusions}

A custom NIM module, based on CAEN A7585D digital power supply, has been
developed to allow an effective online correction of the effect of temperature
on the gain of SiPM arrays. The effect, depending on the type of SiPM array,
is reduced from $30 - 70 \% $ to $\sim 5 \%$.

\end{document}